\documentclass[pre,amsmath,amssymb,aps,twocolumn,superscriptaddress,longbibliography]{revtex4-2}
\usepackage{amssymb}
\usepackage{amsmath}
\usepackage{commath}
\usepackage{verbatim}
\usepackage{graphicx}% Include figure files
\usepackage{dcolumn}% Align table columns on decimal point
\usepackage{bm}% bold math
\usepackage[dvipsnames]{xcolor}
\usepackage[colorlinks]{hyperref}
\usepackage{graphicx}
\usepackage{array}
\usepackage{physics}
\newcolumntype{o}{@{}>{{}}c<{{}}@{}}
\hypersetup{                
	colorlinks=true,                
	breaklinks=true,                
	urlcolor= blue,                
	linkcolor= red,                
    bookmarksopen=false,
	filecolor=black,
	citecolor=blue,
	linkbordercolor=black
}
\begin{document}

%\preprint{APS/123-QED}

\title{Diffusion with two resetting points}% Force line breaks with \\

\author{Pedro Juli\'an-Salgado}
 \email{pjuliansalgado@gmail.com}
\affiliation{Basic Sciences and Engineering, Universidad Autónoma Metropolitana, Apartado Postal 55-534,
Mexico City 09340, Mexico}% 
\author{Leonardo Dagdug}%
 \email{dll@xanum.uam.mx}
\affiliation{Basic Sciences and Engineering, Universidad Autónoma Metropolitana, Apartado Postal 55-534,
Mexico City 09340, Mexico}%
\author{Denis Boyer}
 \email{boyer@fisica.unam.mx}
\affiliation{Instituto de Física, Universidad Nacional Autónoma de México, Mexico City 04510, Mexico}

\date{\today}% It is always \today, today,

\begin{abstract}
We study the problem of a target search by a Brownian particle subject to stochastic resetting to a pair of sites. 
The mean search time is minimized by an optimal resetting rate which does not vary smoothly, in contrast with the well-known single site case, but exhibits a discontinuous transition as the  position of one resetting site is varied while keeping the initial position of the particle fixed, or vice-versa.  
The discontinuity vanishes at a \lq\lq liquid-gas" critical point in position space. This critical point exists provided that the relative weight $m$ of the further site is comprised in the interval $[2.9028...,8.5603...]$. When the initial position is a random variable that follows the resetting point distribution, a discontinuous transition also exists for the optimal rate as the distance between the resetting points is varied, provided that $m$ exceeds the critical value $m_c=6.6008...$
This setup can be mapped onto an intermittent search problem with switching diffusion coefficients and represents a minimal model for the study of distributed resetting.
\end{abstract}

\maketitle

\section{Introduction}
When looking for a lost object or trying to solve a problem,  after some time of unfruitful exploration it might be worth resetting to the starting point and resuming search from there to explore new pathways \cite{evans_diffusion_2011}. Likewise, in a random process whose outcome can be success or failure,  restart can sometimes increase the probability of success \cite{belan2018restart}. The theory of resetting processes has developed considerably in the last decade \cite{evans_stochastic_2020,evans_diffusion_2011-1,kusmierz2014first, kusmierz_optimal_2015,pal_diffusion_2015,campos_phase_2015,bhat_stochastic_2016,pal_diffusion_2016,reuveni_optimal_2016, majumdar_dynamical_2015,evans2018run,pal2019first,nagar2023stochastic} and has allowed to gain understanding in a variety of topics such as the optimization of enzymatic reaction kinetics \cite{reuveni2014role,rotbart2015michaelis,roldan2016stochastic,pal2019landau}, genome evolution \cite{bittihn2017gene,kang2022evolutionary}, computational searches \cite{montanari_optimizing_2002} or animal foraging \cite{boyer2014random,giuggioli2019comparison,pal2020search, viswanathan_physics_2011}.

The paradigmatic resetting process is a Brownian motion (BM) on the line which is stochastically relocated at a constant rate to its starting position \cite{evans_diffusion_2011}. The  mean time needed to reach for the first time a target fixed in space is a non-monotonous function of the resetting rate, with a single minimum at a certain optimal value. This situation is quite generic for processes under resetting \cite{evans2014diffusion}. Resetting is not always beneficial, though:
a general criterion predicts when the mean first passage time (MFPT)  of an arbitrary process at a target state can be decreased or not by stochastic resetting
\cite{reuveni_optimal_2016,pal2017first,chechkin2018random,belan2020median,eliazar2020mean}. Moreover, the fluctuations of the first passage time of resetting processes at optimality share remarkably simple universal properties \cite{reuveni_optimal_2016,belan2020median,riascos2022discrete,bonomo2021first}.

Most of the current knowledge in the field strongly relies on the renewal structure obeyed by any system under resetting to a single state, when the latter also coincides with the initial state (restart). Despite their relevance, the phenomenology of processes where the resetting point can vary from one reset to another is far less known. Distributed resetting was discussed formally for BM \cite{evans_diffusion_2011-1} but studies on this problem and its optimization are scarce 
%and no comprehensive picture has emerged yet 
\cite{
besga_optimal_2020,gonzalez_diffusive_2021,toledo2023first}. With a Gaussian distribution of resetting points, theoretical and experimental results obtained on Brownian micro-spheres confined by optical tweezers have revealed novel features, such as the onset of a local \lq\lq metastable" minimum for the MFPT \cite{besga_optimal_2020}. In the context of search algorithms, random walks on complex networks using stochastic resetting to multiple nodes instead of one can greatly improve their exploration dynamics \cite{gonzalez_diffusive_2021}.

Here we revisit distributed resetting with a systematic study of perhaps the simplest possible case: the two-site problem in the one-dimensional BM. The incorporation of a second resetting point has rather drastic and non-intuitive consequences on the first passage times. In this case,
the distribution of resetting points has both a finite width and a compact support that does not contain the target, in contrast with the case of a single point \cite{evans_diffusion_2011} or points distributed in the entire space \cite{besga_optimal_2020,toledo2023first}. This difference yields new and rich phase diagrams for the optimal rates.
In Section \ref{setup}, we set the problem and provide practical examples. In Section \ref{renewal}, we use a modified renewal approach to calculate the MFPT at a target located at the origin.  Section \ref{results} discusses the behavior of the optimal resetting rate (with an arbitrary initial position of the particle) and unveils abrupt transitions as the parameters of the model are varied. In Section \ref{results2}, similar results are obtained when the initial position follows the resetting point distribution. We conclude in Section \ref{conclusion}.

\section{Setup and motivation}\label{setup}
\begin{figure}
    \includegraphics[width=\linewidth]{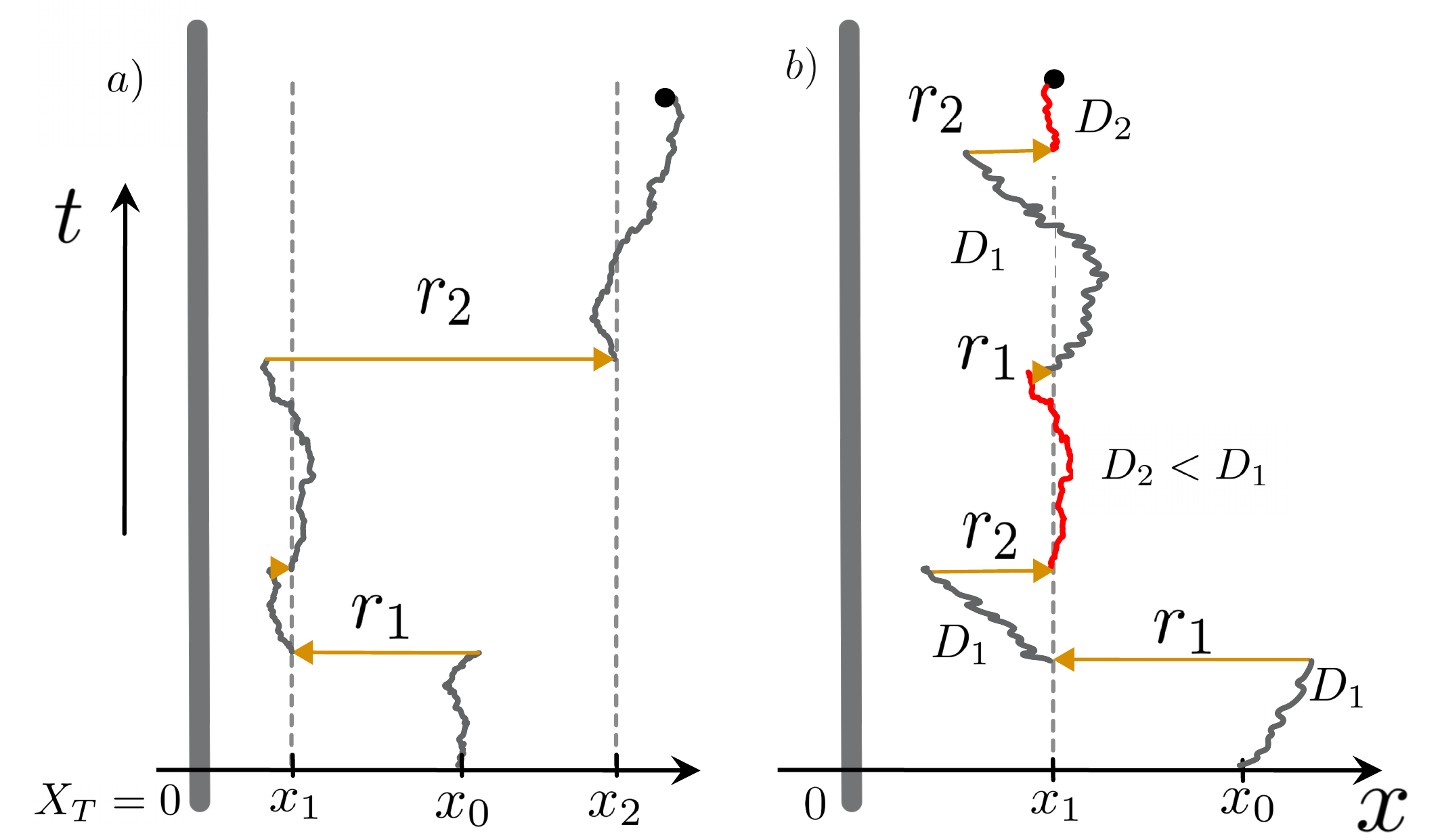}
  \caption{(a) Brownian particle with diffusion coefficient $D$, initial position $x_0$ and resetting rate $r_1$ and $r_2$ to the positions $x_1$ and $x_2>x_1$, respectively. An absorbing boundary is located at $x=0$ on the semi-infinite line. (b) Equivalent intermittent search problem: the particle resets to $x_1$ only with rate $r_T=r_1+r_2$; after each reset, the diffusion coefficient takes two possible values, $D_1=D$ and  $D_2=D(x_1/x_2)^2<D_1$, with probabilities $r_1/r_T$ and $r_2/r_T$, respectively.}\label{fig:ilus_two}
\end{figure}

Let us consider a Brownian particle with diffusion coefficient $D$ and starting point $x_0>0$  on a semi-infinite line with an absorbing boundary at the origin (target). The particle can reset to the fixed positions $x_1$ and $x_2$ with rates $r_1$ and $r_2$, respectively, and we take $x_2>x_1>0$  (see Figure \ref{fig:ilus_two}a for an illutration). Using an infinite line instead of a semi-infinite one would not change the results, due to the symmetry of diffusion and the absorbing target located at the origin.

This model is relevant to several applications. Andean condors ({\it Vultur gryphus}) take refuge in several well-separated roosts, that they use to eat and from where they start new foraging flights \cite{lambertucci2010size,lambertucci2013cliffs}. Individual human mobility patterns are typically dominated by two very frequently occupied places (home and work), while other movements are less predictable \cite{song2010limits}. 

The above configuration can also be interpreted as a search with intermittent dynamics and single-site resetting (Fig. \ref{fig:ilus_two}b). For the first passage times, the problem is equivalent to a diffusion with resetting to $x_1$ only, where the diffusion coefficient can either take a value $D_1=D$ or a smaller value $D_2=D(x_1/x_2)^2$ after each reset. This is due to  the unbounded space to the right and the fact that the diffusion time from $x_2$ to the origin can be written as $\tau_2=x_1^2/(2D_2)$. Models with switching diffusivities have been used to describe gated Brownian ligands in the cell \cite{reingruber2009gated,reingruber2010narrow}, or more generally intermittent searches where a particle can employ two motion modes \cite{benichou_optimal_2005,benichou_intermittent_2011}. In the context of movement ecology, mixtures of random walks with switching dynamics between them are widely used to model or identify discrete foraging states in animals along their trajectory \cite{obrien_search_1990,
kramer_behavioral_2001,morales2004extracting,patterson2008state,fagan2020improved}. In the terminology of \cite{morales2004extracting}, $D_2$ can model a basal \lq\lq encamped" state of local search, while $D_1$ corresponds to an \lq\lq exploratory" motion with larger displacements and possibly incurring higher energetic costs. Here, both modes are interspersed by revisits to a central place ({\it e.g.}, a nest or roost) located at $x_1$.

\section{Renewal approach to the mean first passage time}\label{renewal}

We adopt here a modified renewal equation approach 
%as it constitutes a convenient way to write down the different contributions to the particle survival probability 
\cite{evans_stochastic_2020}. Denoting $S_r\left( t|x_{0},x_{1},x_{2}\right)$ as the probability that the particle under this resetting protocol has not visited the target until time $t$, one has
\begin{multline}
S_{r}\left( t|x_{0},x_{1},x_{2}\right) =S_{0}\left( t|x_{0}\right)
e^{-r_{T}t}\\
+r_{1}\int_{0}^{t }e^{-r_{T}\tau }S_{r}\left( t-\tau
|x_{0}\right) S_{0}\left( \tau |x_{1}\right) \dd{\tau}\\ +r_{2}\int_{0}^{t }
e^{-r_{T}\tau }S_{r}\left( t-\tau |x_{0}\right) S_{0}\left( \tau
|x_{2}\right) \dd{\tau} .  \label{eq1}
\end{multline}
\begin{figure}%[h!]
  \centering
%  \begin{subfigure}{\columnwidth}
    \centering
    \includegraphics[width=\linewidth]{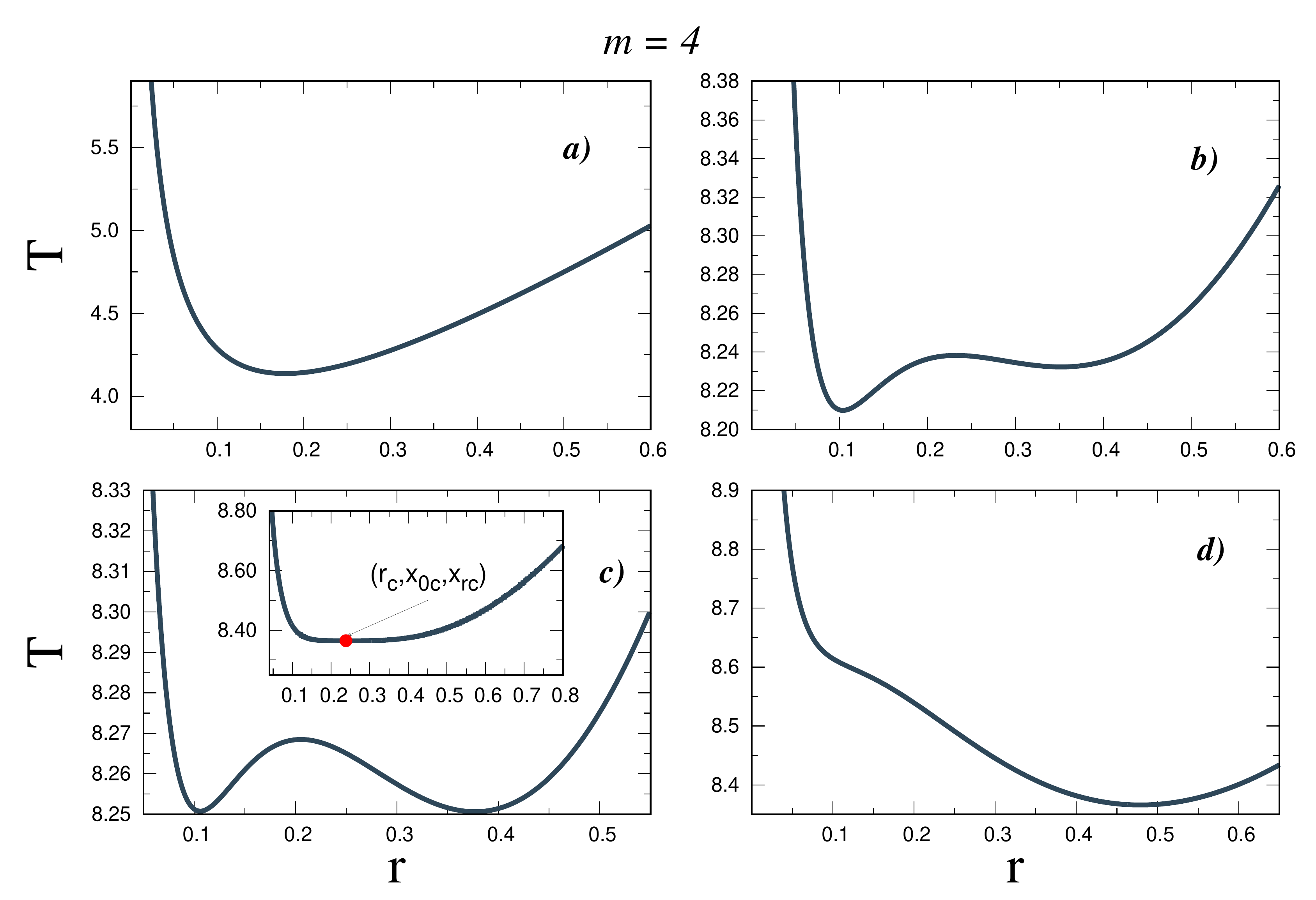}
%  \end{subfigure}%
  \caption{MFPT given by Eq. \eqref{eq7} as a function of $\tilde{r}$, when the position of the further resetting point $\tilde{x}_r$ is increased, fixing $m=4$ and $\tilde{x}_0 = 1.4$. (a) $\tilde{x}_r=2$, (b)  $\tilde{x}_r=3.8$, (c) $\tilde{x}_r=3.82$ ($\simeq$ transition point $\tilde{x}_{rt}$) and (d) $\tilde{x}_r=4$. The optimal rate undergoes a discontinuous transition at $\tilde{x}_{rt}$.
  Inset of (c): for this value of $m$, if $\tilde{x}_0$ is set below but very close to a critical value $\tilde{x}_{0c} =1.64945...$, the two equivalent minima at the transition tend to merge into a single one,
  %(with minimum at $\tilde{r}_c=$), 
  while $\tilde{x}_{rt}$ tends to $\tilde{x}_{rc} = 3.54915...$. For $\tilde{x}_0>\tilde{x}_{0c}$, the transition disappears. }\label{fig:timesVSr}
\end{figure}
Here $r_{T} = r_1 + r_2$ and $S_{0}\left( t|x_{0}\right)$ is the free Brownian particle survival probability given by $S_{0}\left( t|x_{0}\right) = \erf\left( \frac{x_0}{\sqrt{4Dt}}\right)$. The first contribution in Eq. (\ref{eq1}) on the right\textcolor{red}{-}hand side comes from trajectories without resetting in the time interval $[0,t]$, while the last two terms account for resetting to the positions $x_1$ and $x_2$, respectively. The second term describes trajectories whose last resetting event before $t$ occurs at some time $t-\tau$ and at position $x_1$. In the interval $[t-\tau,t]$ the particle diffuses without experiencing any resetting (with probability $e^{-r_{T}\tau}$) and thus survives with probability $S_{0}\left(\tau|x_{1}\right)$. The factor $r_1 d \tau$ stands for the occurrence of a reset to $x_1$ between $t-\tau-d \tau$ and $t-\tau$. The term $S_{r}\left( t-\tau|x_{0},x_{1},x_{2}\right)$ is the probability the particle has not been absorbed in $[0,t-\tau]$. Similarly for the third term of Eq. (\ref{eq1}). 
%Finally the first term on the rhs is the survival probability for the free particle with no resetting events up to time $t$, which occurs with probability $e^{-r_{s}t}$. 
\begin{figure*}
  \centering
%  \begin{subfigure}{\columnwidth
    \includegraphics[width=0.45\linewidth]{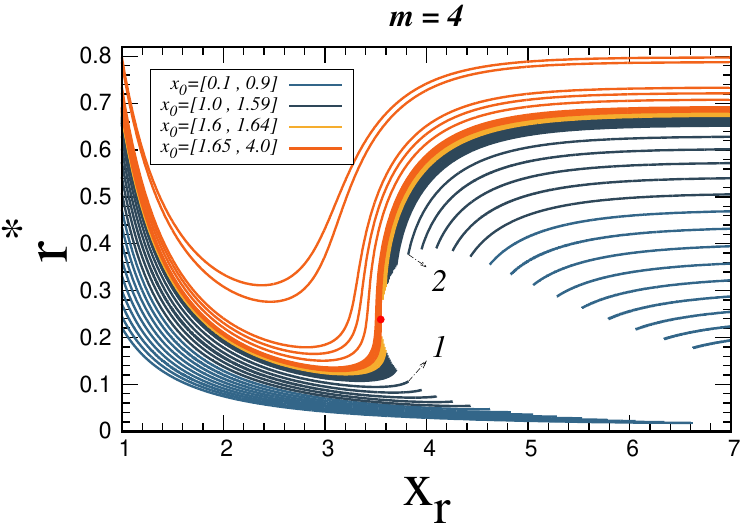}
    \label{fig:jumps_r_T_a}
%  \end{subfigure}%
%  \begin{subfigure}{\columnwidth}
    \includegraphics[width=0.45\linewidth]{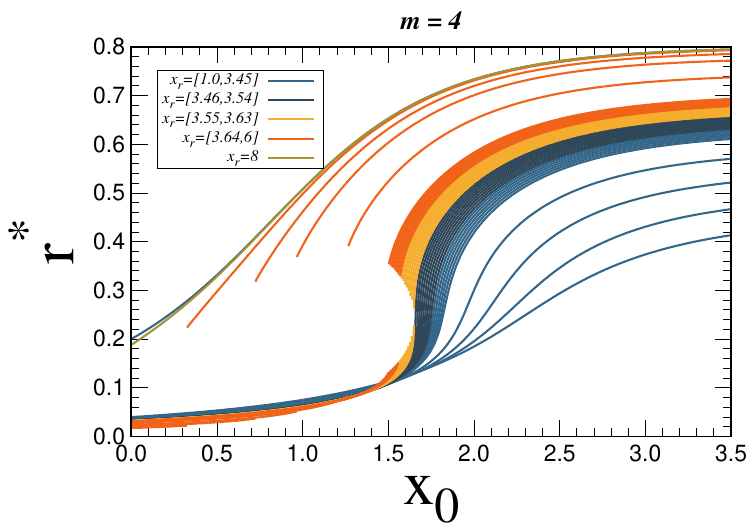}
    \label{fig:jumps_r_T_b}
%  \end{subfigure}%
  \caption{Optimal rate $\tilde{r}$  minimizing the MFPT in Eq. \eqref{eq7} as a function of $\tilde{x}_r$ at fixed $\tilde{x}_0$ (left), or as a function of $\tilde{x}_0$ at fixed $\tilde{x}_r$   (right), fixing $m=4$. The red dot on the left panel indicates the critical point, located at $\tilde{x}_r\simeq 3.5491$, $\tilde{r}^*\simeq 0.23844$ and $\tilde{x}_0\simeq 1.64945$. The numbers 1 and 2 correspond to the minima found in Figure \ref{fig:timesVSr}c at the discontinuous transition ($x_0=1.4$). 
  To avoid excessive notation in the figures, the tildes are not written in the legends.} \label{fig:jumps_r_T}
\end{figure*}

We can reduce Eq. \eqref{eq1} to an algebraic form by taking the Laplace transform ${\cal L}(\cdot)=\int_0^{\infty}e^{-st}(\cdot)dt$ on both sides, leading to
\begin{equation}
   q_{r}\left( s|x_0 , x_1 , x_2 \right) = \frac{q_{0} \left( r_{T} +s |x_0 \right) }{1- r_{1}q_{0} \left( r_{T} +s |x_1 \right) -r_{2}q_{0} \left( r_{T} +s |x_2 \right)}, \label{eq2}
\end{equation}
where $q_r={\cal L}(S_r)$ and $q_0 \left( s | x_i \right) ={\cal L}[S_0(t|x_i)] = \frac{1-e^{-\sqrt{s/D}x_i}}{s}$  (see for example \cite{evans_diffusion_2011}). Therefore
\begin{equation}
    q_{r}\left( s|x_0 , x_1 , x_2 \right) = \frac{1-e^{-\sqrt{\frac{r_{T} + s}{D}}x_0}}{s+r_1e^{-\sqrt{\frac{r_{T} + s}{D}}x_1} + r_2e^{-\sqrt{\frac{r_{T} + s}{D}}x_2}}. \label{eq3}
\end{equation}
As is well known from the theory of first passage processes, 
the mean first passage time is obtained simply by taking $s=0$ in the Laplace transform of the survival probability \cite{bray_persistence_2013}. From Eq. \eqref{eq3}, the MFPT reads
\begin{equation}
  T \left(x_0 ,x_1 , x_2,r_1,r_2 \right)  = \frac{1-e^{-\alpha_0 x_0}}{r_1e^{-\alpha_0 x_1} + r_2e^{-\alpha_0 x_2}}, \label{eq4}
\end{equation}
where $\alpha_{0} = \sqrt{\frac{r_T}{D}}$.  
With $N$ resetting points, the MFPT in \eqref{eq4} would be inversely proportional to $\sum_{i=1} ^{N} r_i e^{-\alpha_0 x_i}$ and $\alpha_0 =\sqrt{\left(r_1 + r_2 + \dots + r_N \right)/D}$.
The expression (\ref{eq4}) reduces to the single point result by taking $r_2=0$ \cite{evans_diffusion_2011}.

For the sake of simplicity in the analysis of expression \eqref{eq4}, we define the dimensionless parameters
\begin{equation}
    m = \frac{r_2}{r_1} , \quad \tilde{r} = \frac{r_1 x_{1}^{2}}{D} , \quad
    \tilde{x}_0 = \frac{x_0}{x_1} \quad \text{and} \quad
    \tilde{x}_r = \frac{x_2}{x_1}, \label{eq5}   
\end{equation}
which is equivalent to set $x_1=1$, $D=1$. Without loss of generality,  
the position $x_1$ is taken the closer to the target, hence $\tilde{x}_r \geq 1$. The parameter $m$ represents the weight of this further resetting point (or the slow diffusion mode $D_2$ in Fig. \ref{fig:ilus_two}b), {\it i.e.}, it is chosen with probability $m/(1+m)$ when any resetting occurs. The re-scaled resetting rate to the closer site is $\tilde{r}$, while $\tilde{r}(1+m)$ is the total rate. We first 
keep $\tilde{x}_0$ as a free parameter to study its impact on optimization, instead of setting $\tilde{x}_0=\tilde{x}_r$ as is usual in the literature.
From Eq. \eqref{eq4}, the dimensionless MFPT reads
\begin{equation}
	  \tilde{T}(\tilde{r},m,\tilde{x}_r,\tilde{x}_0)   = \frac{1-e^{-\tilde{\alpha}_0 \tilde{x}_0}}{\tilde{r}\left( e^{-\tilde{\alpha}_0} +me^{-\tilde{\alpha}_0 \tilde{x}_r}\right)}, \label{eq7}
\end{equation}
with $\tilde{\alpha}_{0}=\sqrt{\tilde{r}\left( 1+m\right) }$ and $\tilde{T}=T/(x_1 ^{2} / D)$. This expression can also be derived following the method of \cite{evans_diffusion_2011-1}.

\section{Optimal resetting rate}\label{results}

\subsection{Behaviour of $\tilde{r}^*$ varying $\tilde{x}_0$ or $\tilde{x}_r$ at fixed $m$}

Figure \ref{fig:timesVSr} displays  
$\tilde{T}$ given by Eq. (\ref{eq7}) as function of $\tilde{r}$ for different positions of the further resetting site, with a fixed initial position $\tilde{x}_0 = 1.4$ and weight $m=4$. In panel (a), the MFPT is non-monotonous and reaches a minimum at an optimal rate, denoted as $\tilde{r}^{\ast}$ in the following.
The MFPT diverges as $\tilde{r}\rightarrow0$ or $\infty$. These properties are similar to the single resetting point case \cite{evans_diffusion_2011}. If $\tilde{x}_r$ increases, however, a new feature appears with a second local,
\lq\lq metastable" minimum at a significantly larger resetting rate [panel (b)].  
When $\tilde{x}_r$ reaches a certain transition value $\tilde{x}_{rt}\simeq 3.82$ [panel (c)], the local and global minima exactly have the same $\tilde{T}$. In other words, there are two well-separated values of $\tilde{r}$ that are optimal.  
A small increase in $\tilde{x}_{r}$ from this point causes the previous local minima to become the global one. Therefore $\tilde{r}^*$ undergoes an abrupt jump to a larger value. For sufficiently large $\tilde{x}_r$ [panel (d)], there is only one minimum again.

The parameter $m$ being fixed, each curve of Figure \ref{fig:jumps_r_T}-left displays the optimal rate for a constant $\tilde{x}_0$. As discussed above, considering $\tilde{x}_r$ as the control parameter, there is a subset of curves which exhibit a discontinuous transition at a value $\tilde{x}_{rt}(x_0,m)$. The points with labels 1 and 2  correspond to the example of Fig. \ref{fig:timesVSr}c with $\tilde{x}_0=1.4$, where the two optimal rates coexist at the discontinuity. Let us denote these rates as $\tilde{r}^{*}_1(\tilde{x}_0,m)$ and $\tilde{r}^{*}_2(\tilde{x}_0,m)$, and  $\Delta \tilde{r}_{12}^{*}(\tilde{x}_0,m)$ their difference. Remarkably, the \lq\lq order parameter" $\Delta \tilde{r}_{12}^{*}$ tends to zero as the initial position $\tilde{x}_0$ approaches a critical value $\tilde{x}_{0c}(m)$, which only depends on $m$. Concomitantly, $\tilde{x}_{rt}(x_0,m)$ tends to a critical value $\tilde{x}_{rc}(m)$, while $\tilde{r}^{\ast}_{1}$ and $\tilde{r}^{\ast}_{2}$ tend to $\tilde{r}_c(m)$. For the \lq\lq isotherms" $\tilde{x}_0>\tilde{x}_{0c}$ (orange curves of Fig. \ref{fig:jumps_r_T}-left), $\tilde{T}$ has a single minimum and the variations of $\tilde{r}^{\ast}$ with $\tilde{x}_r$ are smooth. This scenario recalls the classical $(P,V)$ diagrams of liquid-gas transitions.
Figure \ref{fig:jumps_r_T}-right shows the corresponding diagram obtained by varying $\tilde{x}_0$ at fixed $\tilde{x}_r$, which is quite similar.

The abrupt jump of $\tilde{r}^*$ to a larger value at $\tilde{x}_r=\tilde{x}_{rt}$ can be qualitatively explained as follows. A crucial feature of the two-site problem is the non-monotonous behavior of $\tilde{r}^{\ast}$ with respect to $\tilde{x}_r$ (see Fig. \ref{fig:jumps_r_T}-left and Fig. \ref{fig:Tav} further). For $\tilde{x}_r\sim 1$, the system is close to the one-site problem. If $m\gg1$, most of the resetting occurs to the further site, hence when the latter is taken away from the target, the optimal resetting rate should decrease, roughly as $1/(m\tilde{x}_r^2)$ like in the one-site problem \cite{evans_diffusion_2011}.
However, for too large a $\tilde{x}_r$, the diffusive paths that start from this second point take too long to reach the target: it is preferable to increase  $\tilde{r}$ (smoothly or discontinuously) in order to start from the closer point more frequently, at a rate that is closer to the optimal rate of this site alone. Meanwhile, the second point becomes much less useful for search.

\subsection{Ginzburg-Landau description of the critical point}

The inset of Fig. \ref{fig:timesVSr}c illustrates how the critical point occurs: near $(\tilde{x}_{0c},\tilde{x}_{rc})$, the two minima of $\tilde{T}$ become symmetrical and tend to merge into a single one, which is located at $\tilde{r}=\tilde{r}_c$. For $m=4$, we find $(\tilde{x}_{0c} , \tilde{x}_{rc}) \simeq (1.64945, 3.54915)$, while $\tilde{r}_c = 0.23844...$.  To calculate these critical parameters for a given $m$, three conditions are required, analogous to those that apply to the Ginzburg-Landau theory of liquid-gas transitions: \emph{(i)} by definition, $\tilde{T}$ 
 is minimum at $\tilde{r}_c$, \emph{(ii)} the concavity of $\tilde{T}$ at $\tilde{r}_c$ changes sign
(as illustrated by Fig. \ref{fig:timesVSr}c compared to its inset),  \emph{(iii)} for $\tilde{x}_0$ smaller than but close to $\tilde{x}_{0c}$, the optimal coexisting rates 
$\tilde{r}^{*}_1(\tilde{x}_0,m)$ and $\tilde{r}^{*}_2(\tilde{x}_0,m)$ lie symmetrically on each side of $\tilde{r}_c$, hence the Taylor expansion of $\tilde{T}$ in powers of $\tilde{r}-\tilde{r}_c$ should not have a cubic term. To sum up,
\begin{equation}
	\left.\frac{\partial^n \tilde{T}}{\partial \tilde{r}^n} \right\vert_{\left(
			\tilde{r},\tilde{x}_{0},\tilde{x}_{r}\right) =\left(
			\tilde{r}_{c},\tilde{x}_{0c},\tilde{x}_{rc}\right) } =0\quad {\rm for}\ n=1,2,3,  
	\label{eq8b}
\end{equation}
with $m$ fixed.
Despite the apparent simplicity of the expression (\ref{eq7}), solving the above system of equations analytically is arduous and a numerical resolution scheme is employed with Mathematica. 

\begin{figure}
  \centering
%  \begin{subfigure}{.5\columnwidth}
    \includegraphics[width=0.9\linewidth]{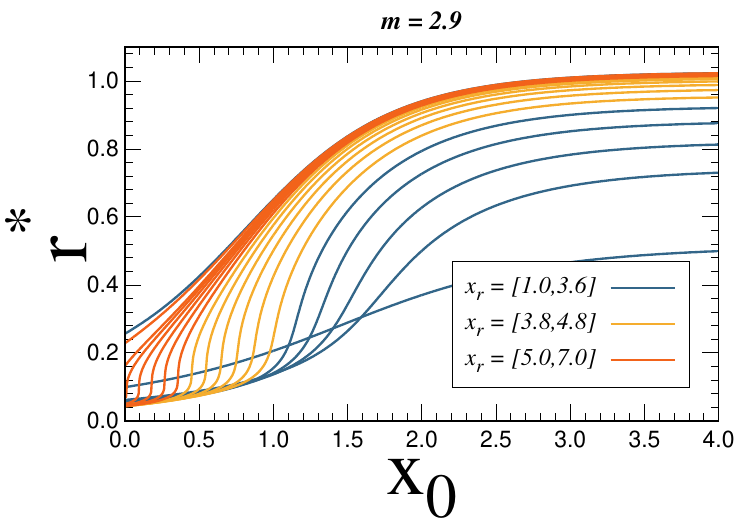}
%  \end{subfigure}%
%  \hfill
%  \begin{subfigure}{.5\columnwidth}
    \centering
    \includegraphics[width=0.9\linewidth]{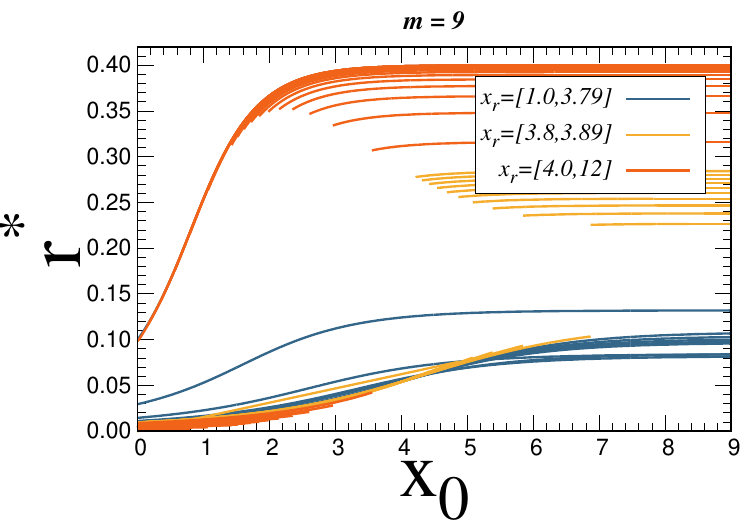}
%  \end{subfigure}%
  \caption{Behaviors of $\tilde{r}^{*}(\tilde{x}_0)$ for two values of $m$ outside of the interval $[m_-,m_+]$, which are quite different from Fig. \ref{fig:jumps_r_T}-right.} \label{fig:m9_m29}
\end{figure}

Generically, the behavior of $\tilde{r}^{\ast}$ close to the critical point in Fig. \ref{fig:jumps_r_T} (left or right) can be described by two scaling laws, for the coexistence and critical curves, respectively. Based on Fig. \ref{fig:jumps_r_T}-left, we have
\begin{eqnarray}
&&\Delta\tilde{r}^{\ast}_{12}\simeq c(\tilde{x}_{0c}-\tilde{x}_0)^{\beta},\quad \tilde{x}_0<\tilde{x}_{0c}\label{scal1}\\
&&(\tilde{r}^{\ast}-\tilde{r}_c)|_{\tilde{x}_0=\tilde{x}_{0c}}\simeq c'|\tilde{x}_r-\tilde{x}_{rc}|^{1/\delta}{\rm sign}(\tilde{x}_r-\tilde{x}_{rc}).
\label{scal2}
\end{eqnarray}
Since the slope of the critical curve is vertical at $\tilde{x}_{rc}$, or $\partial\tilde{r}^{\ast}/\partial \tilde{x}_r|_{\tilde{x}_{0c},\tilde{x}_{rc}}=\infty$, one has $\delta>1$. 

A picture of this transition can be drawn by expanding the MFPT around the critical parameters,
\begin{equation}
\tilde{T}(\tilde{r},m,\tilde{x}_r,\tilde{x}_0)\simeq T_0+a_1(\tilde{r}-\tilde{r}_c)+
\frac{a_2}{2}(\tilde{r}-\tilde{r}_c)^2+\frac{a_4}{4}(\tilde{r}-\tilde{r}_c)^4
\end{equation}
where
$a_1=-\alpha_1[\tilde{x}_r-\tilde{x}_{rt}(\tilde{x}_0)]$ and $a_2=\alpha_2(\tilde{x}_0-\tilde{x}_{0c})$, with $\alpha_1$, $\alpha_2$ and $a_4$ positive constants (the dependence on $m$ is implicit). The coefficient $a_1$ is $<0$ if $\tilde{x}_r>\tilde{x}_{rt}(\tilde{x}_0)$ because minimization favors a larger $\tilde{r}$ in this case (see Fig. \ref{fig:timesVSr}d). By setting $\tilde{x}_r=\tilde{x}_{rt}(\tilde{x}_0)$ and minimizing $\tilde{T}$ with respect to $\tilde{r}$ one recovers Eq. (\ref{scal1}) with a critical exponent $\beta=1/2$ and $c=2\sqrt{\alpha_2/a_4}$. Likewise, by setting $\tilde{x}_0$ to the critical value $\tilde{x}_{0c}$ and using the definition $\tilde{x}_{rt}(\tilde{x}_{0c})\equiv\tilde{x}_{rc}$, the minimization of $\tilde{T}$ yields Eq. (\ref{scal2}) with an exponent $\delta=3$ and $c'=\sqrt[3]{\alpha_1/a_4}$.

\subsection{Dependence of the critical point on $m$}

We now discuss the variations of the critical point $(\tilde{x}_{0c},\tilde{x}_{rc})$ with respect to the weight $m$ of the further site, which are quite non-trivial. Clearly, when $m\rightarrow0$ or $\infty$, the single site problem is recovered. At finite $m$,
however, the existence of a critical point like in Fig. \ref{fig:jumps_r_T} is not guaranteed. A numerical analysis of the system (\ref{eq8b}) actually reveals that it admits a solution only if $m$ belongs to a certain interval $[m_{-},m_{+}]$, with
\begin{equation}
m_{-}=2.902812...\ {\rm and}\  
m_{+}=8.560372...
\end{equation}

\begin{figure}
  \centering
\includegraphics[width=0.9\linewidth]{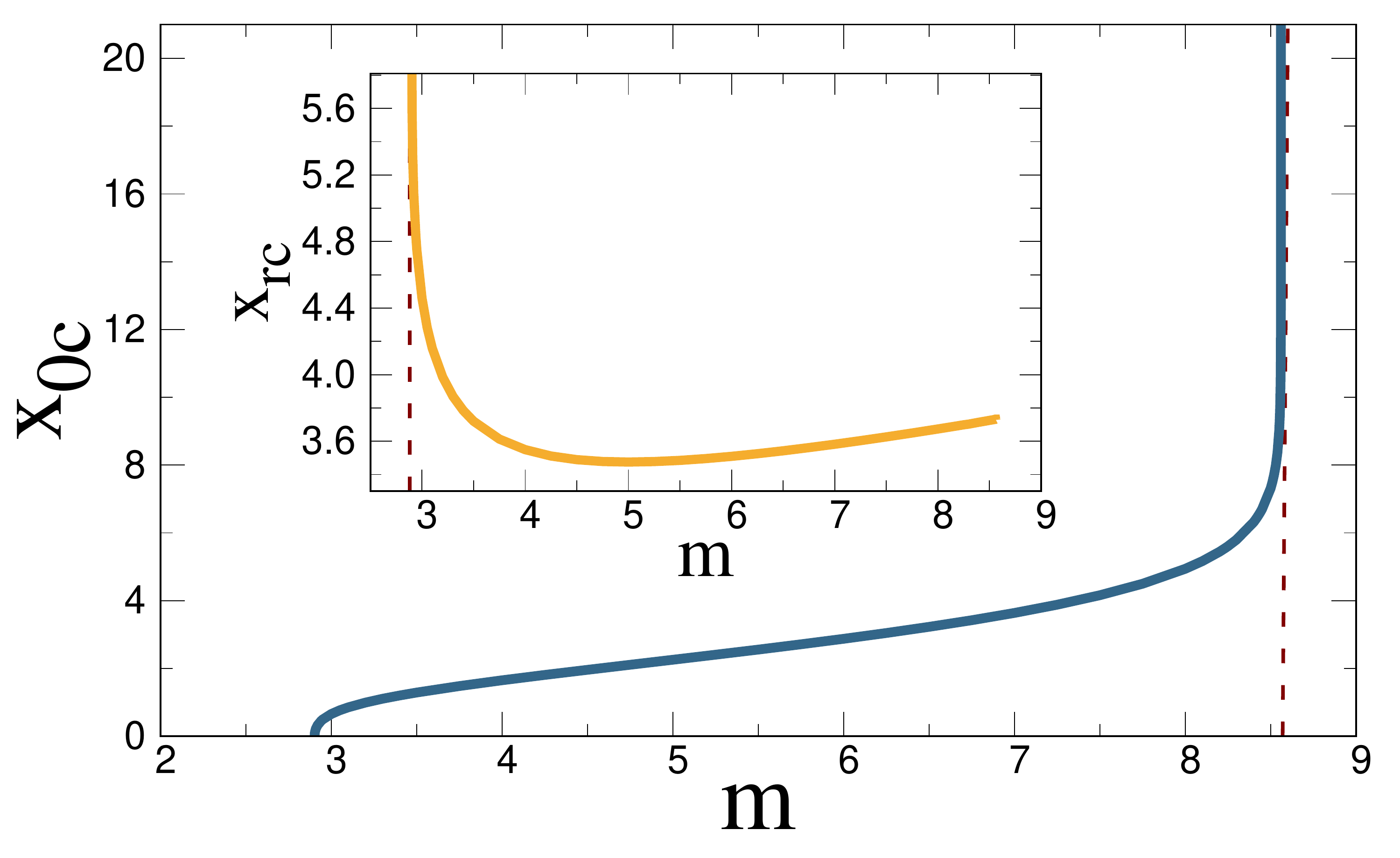}
%  \end{subfigure}%
  \caption{Critical coordinates $\tilde{x}_{0c}$ and $\tilde{x}_{rc}$ (inset) versus $m$.}\label{fig:xsVSm}
\end{figure}

If $m<m_{-}$, the variations of $\tilde{r}^{*}(\tilde{x}_0,\tilde{x}_r)$ are continuous and smooth: in 
Fig. \ref{fig:m9_m29}-top with $m=2.9$, $\tilde{r}^{*}$ monotonically increases with $\tilde{x}_0$ without singularity, the same qualitative behavior as in the single-point case ($m=0$). For $m=9>m_{+}$ (Fig.  \ref{fig:m9_m29}-bottom), $\tilde{r}^{*}$ also increases continuously if $\tilde{x}_r$ is very large or close to $1$. Otherwise, a jump can occur at a certain $\tilde{x}_{0}$, but no critical curve exists. Two coexistence lines describe these discontinuous transitions (the lower one is less clearly visible due to curve overlap) without meeting at a critical point, a situation reminiscent of the liquid-solid transition.
Figure \ref{fig:xsVSm} shows the variations of the critical coordinates $\tilde{x}_{0c}(m)$ and $\tilde{x}_{rc}(m)$ when they exist, {\it i.e.}, for $m\in[m_{-},m_+]$. Notably, $\tilde{x}_{0c}$ vanishes at $m_{-}$ and diverges weakly at $m_{+}$. This agrees qualitatively with the behaviors observed in Fig. \ref{fig:m9_m29}.
Conversely, $\tilde{x}_{rc}$ diverges weakly at $m_{-}$ but tends to a finite value (3.7320...) at $m_{+}$.

\section{MFPT averaged over $\tilde{x}_0$}\label{results2}

We now assume that the initial position $\tilde{x}_0$ is a random variable with the same distribution than the resetting point, a configuration which is particularly relevant in experiments \cite{besga_optimal_2020}. Therefore, $\tilde{x}_0=1$ with probability $1/(1+m)$ whereas $\tilde{x}_0=\tilde{x}_r$ with the complementary probability $m/(1+m)$. The MFPT averaged over $\tilde{x}_0$ is denoted as ${\cal T}$. From Eq. (\ref{eq7}), one obtains
\begin{equation}\label{avmfpt}
{\cal T}(\tilde{r},m,\tilde{x}_r)=\frac{1}{\tilde{r}(e^{-\tilde{\alpha}_0}+me^{-\tilde{\alpha}_0\tilde{x}_r})}-\frac{1}{\tilde{r}(1+m)},
\end{equation}
with again $\tilde{\alpha}_{0}=\sqrt{\tilde{r}\left( 1+m\right)}$. 

\begin{figure}
  \centering
\includegraphics[width=\linewidth]{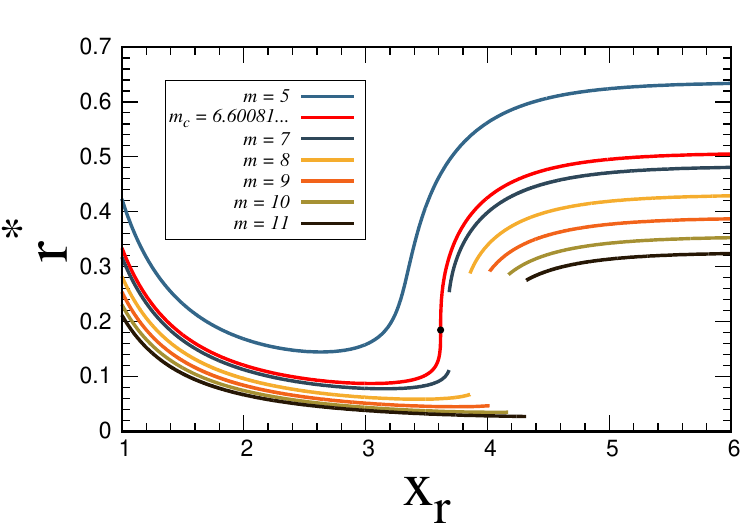}
%  \end{subfigure}%
  \caption{Resetting rate optimizing the $\tilde{x}_0$-averaged MPFT (${\cal T}$) as a function of the position of the further site $\tilde{x}_r$, for several fixed $m$. The black dot indicates the critical point at $(x_c,r_c)$.}\label{fig:Tav}
\end{figure}

The rate $\tilde{r}$ that minimizes ${\cal T}$ is displayed in Fig. \ref{fig:Tav} as a function of $\tilde{x}_r$, for a few representative values of $m$. Similarly to Fig. \ref{fig:jumps_r_T}-left, there are two types of situation. If $m$ lies below a certain critical value $m_c$, the optimal rate $\tilde{r}^*$ has smooth, non-monotonous variations. For $m>m_c$, however, the quantity $\tilde{r}^*$ additionally undergoes a discontinuous jump at a certain \lq\lq coexistence" point $\tilde{x}_{r}=\tilde{x}_{rt}(m)$. In this case, ${\cal T}$ exhibits a metastable behavior with respect to $\tilde{r}$ that is qualitatively the same as the one illustrated by Fig. \ref{fig:timesVSr} for $\tilde{T}$.

The critical weight $m_c$ a well as the critical point $(x_c,r_c)$ where the slope is vertical, represented by a black dot in Fig. \ref{fig:Tav}, can be obtained by following the same steps that led to Eq. (\ref{eq8b}). The critical parameters solve the system of equations:
\begin{equation}\label{criteq2}
\left.\frac{\partial^n {\cal T}}{\partial \tilde{r}^n} \right\vert_{\left(
			\tilde{r},m,\tilde{x}_{r}\right) =\left(
			r_{c},m_{c},x_{c}\right) } =0\quad {\rm for}\ n=1,2,3,  
\end{equation}
and they are readily obtained numerically as
\begin{eqnarray}
 m_c&=&6.600817....\\
 x_c&=&3.615809...\\
 r_c&=&0.184429...
\end{eqnarray}
Likewise, the scaling laws near the critical point now read
\begin{eqnarray}
&&\Delta\tilde{r}^{\ast}_{12}\sim (m-m_c)^{\beta},\quad m>m_{c}\label{scal3}\\
&&(\tilde{r}^{\ast}-r_c)|_{m=m_{c}}\sim |\tilde{x}_r-x_{c}|^{1/\delta}{\rm sign}(\tilde{x}_r-x_{c}).
\label{scal4}
\end{eqnarray}
The critical exponents are unchanged, given by $\beta=1/2$ and $\delta=3$.

\section{Conclusion}\label{conclusion}

We have shown that the addition of a second point to the classic problem of diffusion with resetting \cite{evans_diffusion_2011} gives rise to a wealth of intriguing phenomena which can be studied analytically and recall thermodynamic transitions. Although
this setup had not been systematically analyzed so far, it is rather generic and could be implemented experimentally with optical tweezers: It combines some qualitative aspects of the one-site problem, such as the existence of a finite optimal resetting rate, with aspects of resetting to a broad distribution of points ({\it e.g.} Gaussian), 
where a second local minimum of the MFPT appears as the variance is decreased below a certain value, although no abrupt transition occurs in taht case \cite{besga_optimal_2020}. As show by Fig.  \ref{fig:Tav}, here, it is possible to observe a discontinuous \lq\lq first-order" transition in the optimal rate as the variance (or the distance between the two resetting points) is varied. The existence of this transition is not guaranteed, though, but depends on the shape of the distribution itself  (here, whether the parameter $m$ is larger than a critical value or not). 

When the resetting points are distributed on the whole axis with density $p(z)$, the results would be qualitatively different than those presented here. This is because the density at the target position is not vanishing in general: if $p(z=0)>0$, as for common functions like Gaussians, exponentials or power-laws, the target can be quickly attained by resetting the particle at a very high rate. Therefore $\tilde{r}^{\ast}=\infty$ for all parameter values, irrespective of how small $p(0)$ is 
\cite{besga_optimal_2020}. In contrast,
functions that have a compact support, {\it i.e.}, such that
$p(z)=0$ outside an interval $[a,b]$ which does not contain the target, may display properties similar to the minimal model studied here. A distribution with compact support can describe realistically resetting experiments that trap micro-spheres in a  region of space with optical tweezers, and where the target is not illuminated by the laser beam.

Non-unique minima of the mean search time with abrupt transitions for the optimal parameter have been recently found in other contexts, for instance in the quantum version of the random walk on the line under periodic resetting to a single state. This phenomenon is related in this case to the presence of multiple maxima and minima of the first detection time probability of the reset-free process \cite{yin2023restart,yin2023instability}.

Our results call for more studies on distributed resetting processes and their optimization.  
Discontinuous transitions in the optimal rates are likely to be ubiquitous and
other distributions than the family considered here may lead to a similar phenomenology, or to new features yet to be explored.\\

D.B. and L.D. acknowledge support from Ciencia de Frontera 2019 Grants 10872 and 51476 (Conacyt, México). D.B. thanks Emily Shepard for fruitful discussions.

%Extension: resetting to one of the site depends on the position occupied.

%\bibliographystyle{apsrev4-2}
%\bibliography{biblio}% Produces the bibliography via BibTeX.


\end{document}